\documentclass[10pt, conference, letterpaper]{IEEEtran}

\usepackage[textwidth=1.5in,shadow,backgroundcolor=yellow,textsize=tiny]{todonotes}

\usepackage{mathtools}

\DeclarePairedDelimiter\floor{\lfloor}{\rfloor}

\usepackage[compress]{cite}
\usepackage{microtype} 
\usepackage{amsmath}
\usepackage{amsthm}
\usepackage{tabto}
\usepackage{algpseudocode}
\usepackage{xspace}
\usepackage{paralist}
\usepackage[hyphens]{url}
\usepackage{hyperref}
\usepackage[hyphenbreaks]{breakurl}
\usepackage{subfloat}
\usepackage[caption=false]{subfig}
\usepackage[utf8]{inputenc}
\usepackage{wesa}
\usepackage[T1]{fontenc}

\usepackage{amssymb}
\usepackage{graphicx}
\usepackage{tabularx}
\usepackage{array}
\usepackage{hhline}
\usepackage{multirow}
\usepackage{epsf}
\usepackage{epstopdf}
\usepackage{algpseudocode}
\usepackage[Algorithm]{algorithm}
\usepackage{color, colortbl}
\usepackage{subfiles}
\usepackage{enumitem}

\usepackage{tikz, pgfplots, pgfplotstable}
\usepackage[percent]{overpic} 
\usetikzlibrary{automata,positioning,patterns}
\pgfplotsset{compat=1.15}
\usepgfplotslibrary{units, groupplots}

\newcommand{\ignore}[1]{}

\algnewcommand{\AND}{\textbf{and}\xspace}
\algnewcommand{\OR}{\textbf{or}\xspace}

\newcommand{\abs}[1]{\left\vert#1\right\vert}
\newcommand{\set}[1]{\left\{#1\right\}}

\makeatother
\newbox\statebox
\newcommand{\myState}[1]{%
    \setbox\statebox=\vbox{#1}%
    \edef\thealgruleheight{\dimexpr \the\ht\statebox+1pt\relax}%
    \edef\thealgruledepth{\dimexpr \the\dp\statebox+1pt\relax}%
    \ifdim\thealgruleheight<.75\baselineskip
        \def\thealgruleheight{\dimexpr .75\baselineskip+1pt\relax}%
    \fi
    \ifdim\thealgruledepth<.25\baselineskip
        \def\thealgruledepth{\dimexpr .25\baselineskip+1pt\relax}%
    \fi
    \State #1%
    \def\thealgruleheight{\dimexpr .75\baselineskip+1pt\relax}%
    \def\thealgruledepth{\dimexpr .25\baselineskip+1pt\relax}%
}

\makeatletter
\def\blfootnote{\xdef\@thefnmark{}\@footnotetext}
\makeatother

\newcommand{\missp}{M} 

\newcommand{\ind}{I} 
\newcommand{\indSize}{\abs{\ind}} 
\newcommand{\indSizeT}{\abs{\ind_t}} 
\newcommand{\indSizeTpp}{\abs{\ind_{t+1}}} 
\newcommand{\minIndSize}{\abs {\ind_{\min}}}
\newcommand{\maxIndSize}{\abs {\ind_{\max}}}

\newcommand{\fpp}{FP}
\newcommand{\fnp}{FN}
\newcommand{\updateinterval}{u}
\newcommand{\updateintervalT}{u_t}

\newcommand{\mS} {S}   
\newcommand{\budget}{B}
\newcommand{\bandwidth}{BW}
\newcommand{\cacheSize}{C}
\newcommand{\Meg}{\text{M}}

\newcommand{\alg}{CAB} 

\newcommand{\cachefirst}{CacheFirst}

\newcommand{\minUpdateInterval}{\updateinterval_{\min}}
\newcommand{\maxUpdateInterval}{\updateinterval_{\max}}
\newcommand{\reconfParam}{\alpha}

\newcommand{\plotwidth}{0.8*\columnwidth}

\begin{document}
\title {Self-adjusting Advertisement of Cache Indicators with Bandwidth Constraints}
\author{
 \IEEEauthorblockN
 {
 Itamar Cohen\IEEEauthorrefmark{1},
 Gil Einziger\IEEEauthorrefmark{2},
 and
Gabriel Scalosub\IEEEauthorrefmark{3}
 }
 \IEEEauthorblockA
 {
 \IEEEauthorblockA{\IEEEauthorrefmark{1}
 Department of Electronics and Telecommunications, Politecnico di Torino, Italy. Email: itamar.cohen@polito.it}
\IEEEauthorblockA{\IEEEauthorrefmark{2}Department of Computer Science, Ben-Gurion University of the Negev, Israel. Email: gilein@bgu.ac.il}
 \IEEEauthorblockA{\IEEEauthorrefmark{3}School of Electrical and Computer Engineering, Ben-Gurion University of the Negev, Israel. Email: sgabriel@bgu.ac.il}
 }
}

\maketitle


\begin{abstract}
Cache advertisements reduce the access cost by allowing users to skip the cache when it does not contain their datum. Such advertisements are used in multiple networked domains such as 5G networks, wide area networks, and information-centric networking. The selection of an advertisement strategy exposes a trade-off between the access cost and bandwidth consumption. Still, existing works mostly apply a trial-and-error approach for selecting the best strategy, as the rigorous foundations required for optimizing such decisions is lacking. 

Our work shows that the desired advertisement policy depends on numerous parameters such as the cache policy, the workload, the cache size, and the available bandwidth. In particular, we show that there is no ideal single configuration. Therefore, we design an adaptive, self-adjusting algorithm that periodically selects an advertisement policy. Our algorithm does not require any prior information about the cache policy, cache size, or workload, and does not require any apriori configuration. Through extensive simulations, using several state-of-the-art cache policies, and real workloads, we show that our approach attains a similar cost to that of the best static configuration (which is only identified in retrospect) in each case.

\end{abstract}

\section{Introduction} \label{sec:Intro}
\blfootnote{* The work was done while this author was with Ben-Gurion University.}
Caching is a fundamental optimization technique where a small subset of the data is stored in a cache, which is cheaper to access than the regular storage.
Caching is common to the point where it is present in some form in almost all computing environments and systems, ranging from micro-controllers, through PCs and servers, and onto distributed cloud services.

In large distributed systems, caches often further optimize performance by advertising their content.
Such advertisements allow clients to bypass the cache when it is unlikely to contain the requested datum, thus reducing the total access cost, where ``cost'' can reflect bandwidth, access time, or energy~\cite{BloomParadox, Joint_opt, CDN_theory_Vs_practice}. Content advertisements are  used in mobile ad-hoc networks~\cite{uIntervalInMANET, Digest_in_Manet}, content delivery networks (CDN)~\cite{CDN_OceanStore,CDN_AdaptSize, CDN_theory_Vs_practice}, information centric networking (ICN) ~\cite{ICN,ICN2}, and in wide-area networks~\cite{summary_cache}.

Ideally, the advertisement policy would reflect the cached content at any given time, but such a solution is bandwidth-intensive. Content advertisement is often restricted to a bandwidth budget.
Therefore, systems often compromise on advertising approximate {\em indicators} that approximate the cached content~\cite{Bloom,CBF,TinySet} to reduce the advertisement size, at the cost of some probability of generating false-positive errors~\cite{Bloom,Survey12,Survey18, TinySet,TinyTable,BloomParadox,CDN_OceanStore,AccessEfficientBF}.  
Such errors imply that indicators sometimes mistakenly assert that a datum is stored in the cache, causing redundant cache accesses.  
\begin{figure}[t!]

 

    \begin{tikzpicture}
        log bases x={2},
        log bases y={10},
        \begin{groupplot}[
            group style=
                {
                columns=2,
                xlabels at=edge bottom,
                ylabels at=edge left,
                horizontal sep=0.07\columnwidth,
                group name=plots
                },
    		width  = 0.55*\columnwidth,
     		height = 0.4*\columnwidth,
    		xmode = log,
    		ymode = log,
    		xmin = 2,
     		xmax = 8192,
     		xtick = {2, 16, 128, 1024, 8192},
     		xticklabels = {2,16,128,1K,8K},
     		ymin = 0.0001,
     		ymax = 0.3,
     		ytick = {0.0001,0.001,0.01,0.1},
     		minor tick style={draw=none},
      		yticklabels = {$10^{-4}$,$10^{-3}$,$10^{-2}$, $10^{-1}$},
            ymajorgrids = true,
            ylabel = {Ratio},
     		ylabel near ticks,
    		legend style = {at={(-0.1,1.65)},anchor=north,legend columns=-1,font=\footnotesize},
    		xlabel = {Update interval (\#requests)}, 
    		xlabel near ticks,
    	    label style={font=\footnotesize},
    	    tick label style={font=\footnotesize},
    	]
        	
        \nextgroupplot[
            title = False negative,
            title style ={
                font=\footnotesize,
                yshift = -2pt
                },
     		]
            \addplot[color=black, 	mark=triangle, width = \plotwidth] coordinates {
            (1.0000, 0.0001)(2.0000, 0.0001)(4.0000, 0.0002)(8.0000, 0.0007)(16.0000, 0.0046)(32.0000, 0.0131)(64.0000, 0.0228)(128.0000, 0.0311)(256.0000, 0.0396)(512.0000, 0.0495)(1024.0000, 0.0622)(2048.0000, 0.0781)(4096.0000, 0.0973)(8192.0000, 0.1139)
            };
            
            \addplot[color=red, 	 	mark=o, 				width = \plotwidth] coordinates {
            (1.0000, 0.0001)(2.0000, 0.0001)(4.0000, 0.0002)(8.0000, 0.0010)(16.0000, 0.0065)(32.0000, 0.0184)(64.0000, 0.0320)(128.0000, 0.0438)(256.0000, 0.0558)(512.0000, 0.0697)(1024.0000, 0.0877)(2048.0000, 0.1101)(4096.0000, 0.1367)(8192.0000, 0.1598)
            };
            
            \addplot[color=blue,  	mark=x, 				width = \plotwidth] coordinates {
            (1.0000, 0.0001)(2.0000, 0.0001)(4.0000, 0.0002)(8.0000, 0.0011)(16.0000, 0.0075)(32.0000, 0.0211)(64.0000, 0.0368)(128.0000, 0.0502)(256.0000, 0.0641)(512.0000, 0.0800)(1024.0000, 0.1006)(2048.0000, 0.1262)(4096.0000, 0.1568)(8192.0000, 0.1830)
            };
            
            \addplot[color=purple, mark=+, 				width = \plotwidth] coordinates {
            (1.0000, 0.0001)(2.0000, 0.0002)(4.0000, 0.0003)(8.0000, 0.0012)(16.0000, 0.0077)(32.0000, 0.0216)(64.0000, 0.0376)(128.0000, 0.0513)(256.0000, 0.0655)(512.0000, 0.0818)(1024.0000, 0.1028)(2048.0000, 0.1289)(4096.0000, 0.1602)(8192.0000, 0.1869)
            };

        \nextgroupplot[
            title = False positive,
            title style ={
                font=\footnotesize,
                yshift = -2pt,
                },
      		yticklabels = \empty
      	]
            \addplot[color=black, 	mark=triangle, width = \plotwidth] coordinates {
            (1.0000, 0.2656)(2.0000, 0.2656)(4.0000, 0.2656)(8.0000, 0.2656)(16.0000, 0.2656)(32.0000, 0.2656)(64.0000, 0.2656)(128.0000, 0.2657)(256.0000, 0.2658)(512.0000, 0.2660)(1024.0000, 0.2664)(2048.0000, 0.2673)(4096.0000, 0.2690)(8192.0000, 0.2723)
            };\addlegendentry {2-BPE}
            
            \addplot[color=red, 	 	mark=o, 				width = \plotwidth] coordinates {
            (1.0000, 0.0991)(2.0000, 0.0991)(4.0000, 0.0991)(8.0000, 0.0991)(16.0000, 0.0991)(32.0000, 0.0991)(64.0000, 0.0992)(128.0000, 0.0993)(256.0000, 0.0994)(512.0000, 0.0997)(1024.0000, 0.1003)(2048.0000, 0.1015)(4096.0000, 0.1039)(8192.0000, 0.1084)
            };\addlegendentry {4-BPE}
            
            \addplot[color=blue,  	mark=x, 				width = \plotwidth] coordinates {
            (1.0000, 0.0146)(2.0000, 0.0146)(4.0000, 0.0146)(8.0000, 0.0146)(16.0000, 0.0146)(32.0000, 0.0147)(64.0000, 0.0147)(128.0000, 0.0148)(256.0000, 0.0150)(512.0000, 0.0153)(1024.0000, 0.0160)(2048.0000, 0.0174)(4096.0000, 0.0203)(8192.0000, 0.0256)
            };\addlegendentry {8-BPE}
            
            \addplot[color=purple, mark=+, 				width = \plotwidth] coordinates {
            (1.0000, 0.0003)(2.0000, 0.0003)(4.0000, 0.0003)(8.0000, 0.0003)(16.0000, 0.0003)(32.0000, 0.0003)(64.0000, 0.0004)(128.0000, 0.0005)(256.0000, 0.0007)(512.0000, 0.0011)(1024.0000, 0.0018)(2048.0000, 0.0032)(4096.0000, 0.0061)(8192.0000, 0.0116)
            };\addlegendentry {16-BPE}
        \end{groupplot}3 
    \end{tikzpicture}
    \caption{
    Effect of the update interval on the false-negative errors (left) and false-positive errors (right) for optimally configured Bloom filter indicators. 
    Both axis are in log-scale, and the cache size is 8K, policy is LRU, trace is F1 (described in Sec.~\ref{sec:evaluation}).}
    \label{fig:uInterval_vs_Fp_Fn} \vspace{-0.3cm}
\end{figure}
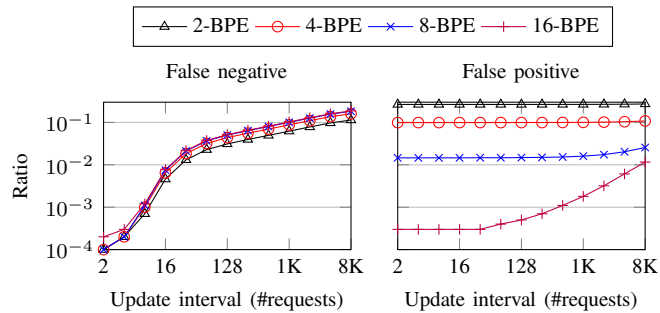

When constantly sending indicator advertisements, the advertised content remains fresh in the sense that it accurately reflects the (approximate) state of the cached content, and we only experience errors due to hash collisions.
In bandwidth-constrained environments, insisting on freshness mandates the usage of relatively small (and inaccurate) indicators to cope with the bandwidth {\em budget}, which would otherwise imply packet drop and increased error rate.
An alternative
is to send an indicator advertisement only {\em occasionally}.
When using such an approach, the advertised content gradually becomes {\em stale}, in the sense that it takes time for the indicator available at the clients to reflect changes in the cached content, which again leads to an increased error rate.

To illustrate the above scenarios, consider an advertisement transmission, followed by having the cache admit a new item ($x$) and evict some item ($y$), such that these events are not (yet) advertised to the clients.
When the client tests for $y$, the indicator falsely indicates that $y$ is in the cache, resulting in a {\em false-positive} error. Similarly, a query for $x$ is likely to falsely indicate that $x$ is not in the cache (as it wasn't in the cache at the time the advertisement was sent), resulting in a {\em false-negative} error.
Thus, staleness creates both false-positive and false-negative errors, and we expect fewer of these errors, the more frequently we refresh the advertisements. 

Fig.~\ref{fig:uInterval_vs_Fp_Fn} shows the percentage of false indications as a function of the time between subsequent advertisements, referred to as the {\em update interval}.
Here, the advertisement size is expressed by the number of {\em Bits Per cached Element} (BPE), while the update interval is expressed in terms of requests between subsequent updates.
Notice that the X-axis and the Y-axis are in logarithmic scale.
Fig.~\ref{fig:uInterval_vs_Fp_Fn} (left) shows that the false-negative ratio increases with the update interval, as we qualitatively explained above, and that it may reach non-negligible rates.
Fig.~\ref{fig:uInterval_vs_Fp_Fn} (right) demonstrates that increasing the update interval also affects the false-positive errors, but the effect is less pronounced than it is on false-negatives.

The combination of approximate indicators and staleness makes selecting a cache advertisement policy a challenging task.
Intuitively, the cache may send an advertisement of size $x$ once every $y$ requests, or a more accurate advertisement of size $\beta\cdot x$ ($\beta>1$) once every $\beta \cdot y$ requests.
Both options require
similar bandwidth, but it is unclear which of them would do better.
Note that although the probability of false-positive due to hash collisions is relatively well understood~\cite{Bloom}, the errors caused by stale advertisements are difficult to predict as they depend on the workload, the cache policy, and the cache size. 
Furthermore, existing works do not address the complex interplay between advertisement strategy, indicator size, update interval, and access cost~\cite{summary_cache, updateIntervalWebCache05, uIntervalInMANET, uIntervalSimsInCCN}.
Instead, most works fix an advertisement policy by crude estimations and rules-of-thumb, rather than by optimizing it according to the system being used, and the workload being served~\cite{SquidFAQ, SquidSpec}.
While such an approach may work in some scenarios, changes in either the cache policy, the workload, the cache size, or the budget, may deteriorate performance significantly, as we demonstrate in Sec.~\ref{sec:motivation}.

\textbf{Our contribution:}
Our proposed solutions make advertisements easier to use in caching systems.
Our first contribution is the formulation of the problem and a rigorous study of the problem domain.
We perform a simulation-based study using state-of-the-art cache policies and real workloads, demonstrating that the best advertisement strategy depends on numerous factors such as the cache policy, the cache size, the workload, and the communication budget.
Our study implies that using previously developed approaches requires extensive testing under varying conditions, to optimize the advertisement policy. Worse yet, every change in
the system parameters requires revisiting the previous decisions.
Such optimization is rarely done in practice because it is time-consuming, and more importantly, because some affecting parameters are uncontrollable by the system designers. E.g., the workload may change dynamically, and the cache size may vary between deployments.

In light of this challenge, we suggest an adaptive, self-adjusting, algorithm that periodically updates its advertisement strategy.
Through an extensive simulation study using real workloads, and state of the art cache policies, we show that our algorithm matches the performance of the best static strategy, that {\em can only be determined in retrospect}.
Adopting our proposed solution implies that system designers are no longer required to optimize their advertisement policy. The algorithm adapts to the system configuration, as well as to the dynamic workload characteristics in runtime.

\section{Related work}\label{sec:related_work}
Indicators are used to periodically advertise the cache content in multiple networking environments, including wide-area networks~\cite{summary_cache}, content delivery networks~\cite{CDN_OceanStore,CDN_AdaptSize, CDN_theory_Vs_practice,Joint_opt}, information centric networking~\cite{ICN,ICN2, indicators_in_NDN19,ICN_survey_15,ICN_survey13}, and wireless networks~\cite{uIntervalInMANET, Digest_in_Manet}.
Indicators make use of randomized hash-based data structures such as Bloom filters~\cite{Bloom,CBF,Dynamic_BF}, and fingerprint hash tables~\cite{TinyTable, TinySet}. 

While conceptually simple, there are numerous challenges in utilizing indicators. For example, the study in~\cite{BloomParadox} shows that indicators may degrade the performance in some scenarios in what is refer to as the ``Bloom Paradox''. 
The works~\cite{Accs_Strategies_Infocom, chen2020sequential, FN_aware} tackle a distributed scenario where multiple caches send indicators, and the client needs to formulate an access strategy that minimizes the total expected cost.
The work of~\cite{CompressedBF} suggests methods to reduce the transmission overheads of indicators, at the expense of larger local memory consumption. 
The works~\cite{FPfree_Ori, HBA_journal} reduce the transmission overheads by accurately advertising important information, while allowing less important information to be stale, or less accurate.  
The work~\cite{Survey18} surveys many optimizations to indicators, such as the support for removals and dynamic scaling. While such structures are used by our work, the exact construction is not a central part of our work. 

Staleness is a major challenge for cache advertisements. The work of~\cite{fpr_fnr_in_dist_replicas} suggests advertisement strategies that make the impact of staleness on the false-positive ratio and the false-negative ratio predictable. However, their proposed solution requires sending an update whenever sufficiently many bits of the indicator have changed.
This might impose a hefty toll on the bandwidth consumption, and may well violate the available budget. 
The works~\cite{summary_cache, updateIntervalWebCache05, uIntervalInMANET, uIntervalSimsInCCN} perform simulation studies in order to identify ``reasonable'' advertisement policies for some concrete settings (workload, cache size, cache policy, miss penalty). Other works~\cite{b1,b5} address the problem of maintaining a bandwidth budget when sending advertisements using a trial-and-error approach.
Such an approach usually requires a lot of effort on the part of system designers, and as the workload may change, one might still end up exceeding (or under-utilizing) the budget.
In comparison, 
we design a
self-adjusting mechanism that utilizes the budget efficiently without resorting to trial-and-error. 
 
In Squid cache~\cite{SquidSpec}, 
the update interval is fixed and defaults to sending an update once in a hour~\cite{SquidFAQ, SquidSpec}. 
Since such a solution is problematic, Squid's spec defines the problem of scaling the update interval as an "open issue"~\cite{SquidSpec}.
In comparison, our work provides a good solution that adapts the update interval, and the indicator size, to the current situation.  
To the best of our knowledge, our work is the first to automatically optimize the advertisement strategy, while efficiently utilizing a fixed bandwidth budget.

\section{System Model}
\label{sec:Model}

This section formally defines our system model, as well as our notation (which is also summarized in Table~\ref{tbl:notations}).

\begin{table}[t]
	\centering
	\caption{List of Notation}
	\label{tbl:notations}
	\begin{tabular}{|c|p{0.7\columnwidth}|}
		\hline
		Symbol & Meaning \\
		\hline
		\hline
		$\cacheSize$ & Cache size [number of elements] \\
		\hline 
		$\mS_t$ & Set of data items in the cache at time $t$\\
		\hline%
		$\ind_t$ & Indicator at time $t$ \\
		\hline
		$\ind_t(x)$ & Indication for datum $x$ \\
		\hline 
		$\minIndSize, \maxIndSize$ & Minimal, maximal feasible indicator size [bits] \\
		\hline%
		$\fpp_t, \fnp_t$ & False-positive, false-negative estimate of indicator $I_t$ \\
		\hline
        $\missp$ & Miss penalty \\
		\hline 
		$\abs{\ind}$ & Indicator size [bits] \\
		\hline
		$\updateinterval$ & Update interval [number of cache requests] \\
		\hline
		$\minUpdateInterval,\maxUpdateInterval$ & Minimal, maximal update interval [requests] \\
		\hline
		$\budget$ & Bandwidth budget [bits/request] \\
		\hline%
		$T$ & Re-configuration interval [number of cache requests] \\
		\hline%
		$\reconfParam$ & tradeoff parameter balancing accuracy / responsiveness / bandwdith variation \\
		\hline%
		$P$ & Cache policy \\
		\hline
		$W$ & Workload (trace) \\
		\hline
	\end{tabular}
\end{table}

{\bf Cache and cost model:}
We consider a cache that contains, at any time $t$, some set of items $\mS_t$.
The maximal number of items in the cache is $\cacheSize$.
Clients issue a sequence of requests/queries for data. The clients may request the data from the cache, or from some remote storage.
Without loss of generality, we refer to each request as arriving in a unique {\em time slot} $t$. Yet, when clear from the context, we sometimes omit the subscript $t$. 

Accessing the cache incurs some {\em access cost}, which we normalize to 1 without loss of generality.
Cache access cost is due whenever the cache is accessed, even if the requested datum is not in the cache. 
If a cache access for datum $d$ at time $t$ results in a {\em cache miss}, i.e., $d \notin S_t$, then an additional {\em miss penalty} $M>1$ is incurred.
This miss penalty is also imposed whenever the cache is not accessed for a given request.
The miss penalty reflects the cost of retrieving the datum from some remote storage.
The cost $\missp$ includes notifying the cache about the data access, in which case the cache may decide to admit the datum towards serving future requests, depending on the policy being applied for admitting and evicting items from the cache. 
The {\em service cost} is the sum of the cache access cost, and the miss penalty cost.
To make a meaningful comparison of performance, we focus our attention on the {\em average service cost} of all the requests in the sequence, thus following similar cost models studied in previous works~\cite{Accs_Strategies_Infocom, FN_aware, LRUVsFifoRoyOhad}.

{\bf Indicators, update intervals, and configurations:}
At any time $t$, the cache may advertise an {\em indicator} $\ind_t$ that approximates $\mS_t$ at time $t$.
For any indicator $\ind_t$, given a datum $x$, a {\em positive indication} of $\ind_t$ indicates that $x\in \mS_t$, while a {\em negative indication} of $\ind_t$ indicates that $x \notin \mS_t$.
$\ind_t$ may generate false-positive and false-negative errors. 
A positive indication is said to be a {\em false-positive} when $x \notin \mS_t$.
Similarly, a negative indication is said to be a {\em false-negative} when $x \in \mS_t$.
We let $\fpp_t$ and $\fnp_t$ denote the estimates at time $t$ of the false-positive probability, and the false-negative probability, of a cache request, respectively.
We let $\indSizeT$ denote the size of the indicator $\ind_t$ in bits.
To use only feasible sizes, the indicator size should be within some predefined range $[\minIndSize, \maxIndSize]$.

Given some positive integer $T$, we consider a non-overlapping partitioning of time (or equivalently, the sequence of requests) into {\em segments} of length $T$.
The {\em update interval} $\updateintervalT$ is the number of requests between subsequent indicator updates that the cache sends to the users.
At any time $t$, $\updateintervalT$ represents the time between the last update that was sent, and the next update scheduled to be sent.
When considering dynamic algorithms, we allow the value of $\updateintervalT$ to be adjusted only at the end of a segment.

The update interval is at least $\minUpdateInterval$.
One could use $\minUpdateInterval = 1$, but a slightly higher interval enables piggybacking indicator updates on packets carrying cached data payloads, to avoid transmission overheads~\cite{Digest}. We also use a maximal update interval denoted $\maxUpdateInterval$ (which we discuss in the sequel).
We refer to the tuple $(\indSizeT, \updateinterval_t)$ as a {\em configuration}. 

An advertisement that includes the full indicator $\ind$, is called a {\em full-indicator} update. Alternatively, an update that contains the list of bits in the indicator that have flipped since the previous advertisements is called a {\em delta} update.
Specifying the location of each bit in the indicator requires $\log \indSize$ bits. We assume that the cache uses a delta update whenever this consumes less bandwidth than sending a full indicator, namely, when the number of bits flipped in the indicator since the last update is less than $\frac{\indSize}{\log \indSize}$. 

{\bf Bandwidth constraints:}
To model the system's bandwidth constraint, we use the previously defined partitioning of time into segments.
The transmitted {\em bandwidth cost} of configuration $(\indSizeT, \updateinterval_t)$ over a segment of length $T$ ending at time $t$ is the {\em average} number of update bits per request, being {\em sent} to the user during the segment.
We denote this cost by $\bandwidth_t$.
Since indicators are usually of size $\Theta(C)$, and since we would like to potentially allow the algorithm to transmit more than one update during a segment, we require that 
$T \geq \max\set{\maxUpdateInterval,\cacheSize}$.
In particular, we choose 
$T = \reconfParam \cdot \max\set{\maxUpdateInterval,\cacheSize}$, for some positive integer $\reconfParam$.
Parameter $\reconfParam$ serves to control the tradeoff between
\begin{inparaenum}[(i)]
\item the variance of the statistics gathered during a segment, and
\item the dynamic response of the algorithm across segments.
\end{inparaenum}
I.e., if $\reconfParam$ is small, then statistics are gathered over a short interval, and may capture only very transient behavior which could be very different in the following segment. On the other hand, if $\reconfParam$ is large, then the algorithm maintains its current configuration longer, even though workload and system characteristics may change significantly during the segment.

We target system configurations that satisfy budget constraints, defined by a {\em bandwidth budget} of $\budget$ bits/request.
The budget constraint requires that the bandwidth cost in each segment is at most $\budget$. 
We note that when sending a full indicator in each update we must have
\begin{equation}
\label{Eq:budget_const_full_ind}
\frac{\indSizeT}{\updateinterval_t} \leq \budget.
\end{equation}
We use this equation for determining $\maxUpdateInterval$ as the minimal value satisfying Eq.~\ref{Eq:budget_const_full_ind}, which implies that
$\maxUpdateInterval=\floor[\big]{\frac{\maxIndSize}{\budget}}$.

A configuration $(\indSize,\updateinterval)$ is said to be {\em static} if for every time $t$, $\indSizeT=\indSize$ and $\updateinterval_t=\updateinterval$.
Such a configuration is said to satisfy the budget constraint if in {\em every} segment of length $T$, the overall bandwidth cost of using $(\indSize,\updateinterval)$ is at most $\budget\cdot T$, i.e. $\bandwidth_t \leq \budget$ for every time $t$ in which a segment ends.
We note that due to the dynamic nature of caching environments, it may be impossible to verify {\em a-priori} that a specific static configuration does not violate the budget constraint. In particular, a  configurations that uses delta updates might end up violating the budget if there are too many updates.

In our work, we are interested in {\em dynamic} configurations that may re-scale and adjust both $\indSizeT$ and $\updateintervalT$ over time.
Such dynamic configurations may also occasionally end up oversubscribing the network.
However, using dynamic configurations one can strive to satisfy the budget constraint over all segments, by adjusting to the current workload pattern, while (implicitly or explicitly) taking into account additional system parameters related to, e.g., the cache size, or the cache policy.
Although dynamic configurations may sometimes violate the budget constraint, a careful adjustment of the configurations throughout the system's lifetime may reduce this violation significantly (e.g., compared to static configurations).

Since configurations (either static or dynamic) may end up violating the budget constraint, we apply a {\em network policing} mechanism that enforces the budget constraint as follow:
At the beginning of each segment, the cache receives $B \cdot T$ tokens.
Once the overall number of bits sent for indicator advertisement during the segment reaches $B \cdot T$, all further updates during the segment are dropped by the network policing mechanism.
This model conforms to common network policing behaviour that may selectively drop packets when a user oversubscribes its allotted resources.
In this sense, the transmitted bandwidth cost $\bandwidth_t$ may indeed be larger than the budget, but in effect, the network will never forward more traffic than the amount prescribed by the budget $\budget$.
Lastly, we note that $\reconfParam$ also serves to define the time horizon for which we enforce the budget violation.
I.e., choosing a larger value for $\reconfParam$ implies that we allow larger fluctuations in bandwidth usage during a segment, as long as the overall bandwidth cost is maintained over the entire segment.

In what follows, we consider distinct system {\em scenarios}, where each scenario is defined by the cache size $\cacheSize$, policy $P$, workload $W$, and budget $\budget$.
We denote such a scenario by $(\cacheSize,P,W,\budget)$.
We will be studying static advertisement configurations for a variety of scenarios, as well as dynamic advertisement configuration strategies that {\em adapt} to dynamically changing scenarios.
Our work considers the problem of (dynamically) adjusting the configuration so as to minimize the (average) service cost within a  given bandwidth budget.
Throughout our work, all logarithms are of base 2.

\section{Motivation and Preliminaries}
\label{sec:motivation}

This section provides insights into the performance of static configurations to further motivate dynamic advertisement strategies.
We present the results of several experiments, which use several real-life workloads and state-of-the-art cache policies.
Our results show that there is no ``one-size-fits-all'' configuration and that using static configurations may lead to substantial performance degradation in highly dynamic systems.

For each scenario $(C,P,W,B)$ considered, we perform a grid-search of static configurations
and find the best configuration that satisfies the bandwidth budget constraint for this scenario.
We then compare the performance of these configurations when used for scenarios that differ by merely one aspect, where we focus here on changing either the cache size, the policy, or the workload, to exemplify the effect each of these system aspects has on system performance.
In this evaluation, we set $\minUpdateInterval=10$, $\minIndSize=2.5 \cdot C$, $\maxIndSize=15 \cdot C$, $\missp=3$, $\alpha=10$, and $\budget=20$.%
\footnote{We describe the specific workloads and cache policies, as well as the methodology of our grid search and the choice of parameters, in Sec.~\ref{sec:evaluation}.}
We demonstrate configurations that exhibit very good performance for some scenarios, but changing merely one aspect in the scenario results in significant performance degradation.

\setlength\abovecaptionskip{-5pt}

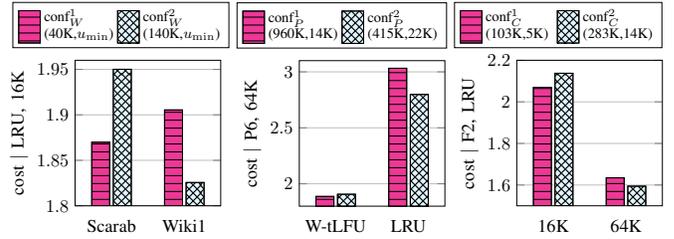
\begin{figure}

\newcommand{\confdiffplot}[9]{

\pgfplotsset{
    /pgfplots/ybar legend/.style={
    /pgfplots/legend image code/.code={%
      \draw[##1,/tikz/.cd,yshift=-1em]
        (0cm,0cm) rectangle (12pt,1.8em);},
  },
}
\begin{tikzpicture}
    \begin{axis}[
        ybar,
        width  = 0.5*\columnwidth,
        height = 0.5*\columnwidth,
        major x tick style = transparent,
        enlarge x limits=0.5,
        ymajorgrids = true,
        ylabel = {#7},
        ylabel shift = -2pt,
        symbolic x coords={
            y1,
            y2,
        },
        xticklabels={
            #3,
            #4,
        },
        xtick=data,
        ymin = #5,
        ymax= #6,
        legend style={
            font=\footnotesize,
            at={(-0.43,1.4)},
            anchor=north west,
            legend columns=2,
            cells={align=left},
        },
    ]
    \addplot[
        fill=magenta!90,
        postaction={pattern=horizontal lines}
        ]
        table [y=conf1]   {#1};
    \addlegendentry{conf$_{#2}^1$\\#8};
    \addplot[
        fill=cyan!10,
        postaction={pattern=crosshatch}
        ]
        table [y=conf2]   {#1};
    \addlegendentry{conf$_{#2}^2$\\#9};
    \end{axis}
    
\end{tikzpicture}
} 



    
    \hspace{-0.3cm}
    \scalebox{.68}{\confdiffplot{motivation_fig_change_trace.dat}{W}{Scarab}{Wiki1}{1.8}{1.96}{cost $|$ LRU, 16K}{(40K,$\minUpdateInterval$)}{(140K,$\minUpdateInterval$)}}
    \hspace{-0.4cm}
    \scalebox{.68}{\confdiffplot{motivation_fig_change_policy_P6.dat}{P}{W-tLFU}{LRU}{1.8}{3.1}{cost $|$ P6, 64K}{(960K,14K)}{(415K,22K)}}
    \hspace{-0.4cm}
    \scalebox{.68}{\confdiffplot{motivation_fig_change_cache_size.dat}{C}{64K}{16K}{1.5}{2.2}{cost $|$ F2, LRU}{(103K,5K)\ \ }{(283K,14K)}}
    \caption{Differences in cost for static configurations when varying workload (left), policy (center), and cache size (right).
    For each aspect being compared, one configuration is better for one value (workload / policy / cache size), whereas another configuration is better for the other value.}
    \label{fig:motivation}
\end{figure}

Fig.~\ref{fig:motivation} shows the results of several such experiments.
In Fig.~\ref{fig:motivation} (left) we consider two workloads $W_1, W_2$, where $W_1$ is the Scarab trace, and $W_2$ is the Wiki1 trace, with conf$^{W}_1$ and conf$^{W}_2$ being their best static configurations, respectively. 
The policy is LRU, and the cache size is $\cacheSize=16$K.
One can note that for each workload $W_i$, using conf$^{W}_j$ $j \neq i$ incurs a toll as large as 5\% compared to using conf$^{W}_i$.
In Fig.~\ref{fig:motivation} (center), one can see similar results hold for varying the cache policies. Here we consider policies $P_1, P_2$, where $P_1$ is W-tLFU and $P_2$ is LRU, with conf$^{P}_1$ and conf$^{P}_2$ being their best static configurations, respectively. The workload is P6, and the cache size is $\cacheSize=64$K.
In Fig.~\ref{fig:motivation} (right), one can see the same effect is manifested for the case where we vary the cache size. Here we consider cache sizes $\cacheSize_1=16K$ and $\cacheSize_2=64K$, with conf$^{\cacheSize}_1$ and conf$^{\cacheSize}_2$ being their best static configurations, respectively. The cache policy, in this case, is LRU, and the workload is F2.
In all plots, the configurations are specified in the legend (up to rounding to the nearest K).

\section{Algorithm \alg}
\label{sec:alg}

In this section, we introduce the Cache-indicators Advertisement with Budget constraint (\alg) algorithm.
The pseudo-code of \alg\ is provided in Algorithm~\ref{alg:alg}.
We begin with a high-level description of our algorithmic concepts and then detail the optimization it employs in its decisions.

\subsection{High-Level Overview}\label{sec:alg:HL_description}
\alg\ has two challenges, in each of two {\em regimes}:
\begin{inparaenum}[(i)]
\item when sending full indicators, and \item when sending delta updates.
\end{inparaenum}

\subsubsection{Full-Indicator Regime}
When sending full indicators, the problem is to find the right balance between the update interval and the accuracy of the indicator.
In this regime, a configuration $(\indSizeT, \updateinterval_t)$ sends an advertisement of size $\indSizeT$ once in $\updateinterval_t$ requests.
Recall that by Eq.~\ref{Eq:budget_const_full_ind}, we must satisfy $\frac{\indSizeT}{\updateinterval_t} \leq \budget$.
Here, our approach is to choose, among all the possible configurations that satisfy Eq.~\ref{Eq:budget_const_full_ind}, a configuration that equalizes the additional cost caused by false-negative errors (controlled by the update interval) and false-positive errors (mainly controlled by the indicator size). 

\subsubsection{Delta Regime}
When only sending the bits that have changed since the previous advertisement, increasing the update interval would usually have very little effect on the consumed bandwidth.
Intuitively, updating about a single change in the cache once every $u$  requests consumes a similar bandwidth as updating about $x$ changes in the cache once in every $x \cdot u$ requests.
The only exception to this rule-of-thumb happens due to hash collisions, e.g., when some of the $x$ changes in the cache occasionally flip and re-flip the same bit in the indicator.
However, when $u$ is small, this effect is negligible.
Hence, we favor sending updates as soon as possible,
i.e., we set the update interval to $\minUpdateInterval$.

Once the update interval is fixed, the remaining challenge is to dynamically-scale the indicator size $\indSize$ to utilize all the budget without exceeding it.
However, sending changes may become infeasible (e.g., when the hit ratio drops and more items are admitted to the cache), and in that case, we might need to return to the full indicator regime. 
We trigger such a transition when we exceed the budget while using the minimal indicator size, $\minIndSize$. 
To do so, we increase the update interval to the ``safe zone'' of sending full indicators, i.e., satisfying Eq.~\ref{Eq:budget_const_full_ind}, in which case we can assure compliance with the budget constraint.
Such a step allows us to search for better configurations in subsequent segments (as also demonstrated in Sec.~\ref{sec:evaluation}).

\usetikzlibrary{fit,calc}
\newcommand*{\tikzmk}[1]{\tikz[remember picture,overlay,] \node (#1) {};\ignorespaces}
\newcommand{\boxit}[5]{\tikz[remember picture,overlay]{\node[yshift=3pt,fill=#1,opacity=.25,fit={($(A)-(#2\linewidth,#3\baselineskip)$)($(B)+(#4\linewidth,#5\baselineskip)$)}] {};}\ignorespaces}
\colorlet{mode1}{red!40}
\colorlet{mode2}{cyan!60}
\colorlet{mode3}{green!70}

\begin{algorithm}[t!]
\caption{\alg ($\budget$)}
\label{alg:alg}
\begin{algorithmic}[1]
    \State $\abs{\ind_0} = \abs {\ind_{\min}}, \updateinterval_0 =  \floor*{\frac{\abs{\ind_0}}{\budget}}$
    \label{alg:init-conf}
    \State $T = \reconfParam \cdot  \max \set { \maxUpdateInterval, \cacheSize }$
    \label{alg:init-T}
    \For {every time slot $t = T, 2T, 3T, \dots$ } 
    \label{alg:mainLoop-start}
        \tikzmk{A}
    	\If {$\exists$ full indicator update during $[t-T,t)$}
    	\label{alg:full_indicator_begin}
        	\State $\abs {\ind_{t+1}} =$   
         	FitToRange $\left(
         	\floor*{ \abs{\ind_t}\sqrt {\frac{\fpp_t}{(\missp-1) \cdot \fnp_t}} }
         	\right)$
         	\label{alg:sqrt_rule}
        	\State $\updateinterval_{t+1} = \floor[\big]{\frac{\indSizeTpp}{\budget}}$
        	\label{alg:full_indicator_end}
        \Else
        \tikzmk{B}
        \boxit{mode1}{0.82}{-0.1}{0.86}{0.65}
        \Comment{all updates are delta-updates}
            \If{$\indSizeT > \minIndSize\ \OR \ \bandwidth_t \leq \budget$} 
            \label{alg:delta_lambert_begin}
            \tikzmk{A}
            \State $\updateinterval_{t+1} = \minUpdateInterval$
        	    \State $\indSizeTpp = $
        	        FitToRange $\left(
        	        \floor[\Big]{
            	    e^{W\left(
            	    \frac{B \indSizeT\log\indSizeT}{\bandwidth_t}
            	    \right)}
            	    } \right)$
        	    \Comment{Lambert $W$ function}
    	        \label{alg:delta_lambert_end}
            \Else
            \tikzmk{B}
            \boxit{mode2}{0.823}{-0.2}{0.798}{0.8}
            \tikzmk{A}
            \Comment{$\indSizeT = \minIndSize\ \AND \ \bandwidth_t > \budget$} 
            \label{alg:delta_min_violation_begin}
            \State $\indSizeTpp = \indSizeT$
            \Comment{indicator size remains $\minIndSize$}
            \State $\updateinterval_{t+1} = \floor*{\frac{\indSizeTpp}{\budget}}$
           \label{alg:delta_min_violation_end}
	        \EndIf
        \EndIf
        \tikzmk{B}
        \boxit{mode3}{0.268}{-0.2}{0.82}{0.8}
    \EndFor
    \label{alg:mainLoop-end}
    \Statex \hrulefill
    \Procedure{\normalfont FitToRange(size)}{}
    \label{alg:FitToRange-start}
        \State \Return $\max \set{ 
    	   \min \set{\textrm {size}, \maxIndSize}, 
    	   \abs {\ind_{\min}}
    	   }$
    \EndProcedure
    \label{alg:FitToRange-end}
\end{algorithmic}
\end{algorithm}

\subsubsection*{Co-similarity and system lifetime}
Our algorithm implicitly assumes that the behaviour of the cache in the next segment will be similar to its behaviour in the current segment. 
While such an assumption is not always correct, many underlying caching algorithms make similar assumptions. E.g., adaptive caches~\cite{Adaptive_Sw_Cache,ARC} assume that the past access pattern provides a good indication of the future access pattern. Thus, our assumptions are reasonable in workloads where adaptive caching works well~\cite{Adaptive_Sw_Cache,ARC,Hyperbolic}.
It should be noted that in Sec.~\ref{sec:evaluation} we demonstrate the effectiveness of our approach in a variety of scenarios and workloads.
\alg\ is oblivious of the cache policy that manages the evictions and admissions, as we show in Sec.~\ref{sec:evaluation}. 
Instead, \alg\ uses only the information of false-positive and false-negative errors, as well as the bandwidth cost, to adapt its advertisement strategy.
Such information indirectly includes some details about the cache policy, e.g., when the cache policy rapidly changes the cached content, then a large update interval is likely to cause plenty of false-negative errors. Alternatively, when the cache policy hardly changes the cached content, the same (long) update interval results in few false-negative errors. 

\subsection{Detailed Description}

The \alg\ algorithm (formally defined in Algorithm~\ref{alg:alg}) is an implementation of the approach outlined above. 
The algorithm begins with an arbitrary configuration that satisfies the budget constraint (line~\ref{alg:init-conf}).
The algorithm then sets the segment size $T$ (line~\ref{alg:init-T}), such that the configuration may be updated at the end of each segment (at times $t=T,2T,\ldots$).
$T$ is set to ensure that sufficient statistics can be obtained during a segment, for determining the configuration to be used towards the following segment.
This is controlled by the value of $\reconfParam$, as described in Sec.~\ref{sec:Model}.
We note that it usually suffices to set $\reconfParam$ to a small constant number (e.g., throughout our evaluation in Sec.~\ref{sec:evaluation} we use $\reconfParam=10$).
At every time $t=T,2T, \ldots$, we let $\fpp_t$ and $\fnp_t$ denote the false-positive ratio and the false-negative ratio during the segment ending at $t$, and we recall that $\bandwidth_t$ denotes the bandwidth cost (i.e., number of bits being sent by \alg, divided by $T$) during this segment.
Whenever a configuration is chosen, the procedure FitToRange (lines~\ref{alg:FitToRange-start}-\ref{alg:FitToRange-end}) ensures that the indicator size is within the prescribed bounds, i.e., in $[\minIndSize,\maxIndSize]$.
For determining  the indicator's size and the update interval, \alg\ distinguishes between three cases (marked by different shaded colors in Algorithm~\ref{alg:alg}). In what follows, we discuss the algorithmic design criteria for each of these cases.

\subsubsection{Full indicator updates}
The first case (lines~\ref{alg:full_indicator_begin}-\ref{alg:full_indicator_end}) is when a full indicator is sent in (at least) one of the updates during the segment ending at $t$.
We refer to this case as having the algorithm work in {\em Mode 1} (shaded red).
In this operation mode, we expect to have $\fpp_t>0$ due to using indicators (which by nature provide merely an approximate representation of the cached content), and $\fnp_t>0$ due to staleness. 
\alg\ adjusts its indicator size and update interval in an attempt to strike a balance between the loss of performance caused by false-negatives, and false-positives.
This approach makes the reasonable assumption that false-negatives increase when increasing the update interval (as exhibited, e.g., in Fig.~\ref{fig:uInterval_vs_Fp_Fn}), and false-positives increase when decreasing the indicator size. Due to the budget constraint, the indicator size and the update interval are  positively correlated.

For understanding the choice made in line~\ref{alg:sqrt_rule}, one should note that
\begin{inparaenum}[(i)]
\item a false-positive indication incurs an unwarranted extra cost of 1, whereas
\item a false-negative indication incurs an unwarranted extra cost of $(\missp-1)$ (since we could have incurred a cost of 1 by merely accessing the cache).
\end{inparaenum}
It follows that targeting having $(\missp-1)$ false-positives (which are relatively cheap) for every single false-negative (which is relatively expensive) would balance the unwarranted extra costs.
I.e., we would like to have $(\missp-1)\cdot\fnp=\fpp$, or equivalently, $\frac{\fpp}{(\missp-1)\cdot\fnp}=1$.
When considering $\fpp_t$ and $\fnp_t$, if $(\missp-1)\cdot\fnp_t < \fpp_t$, we would like to decrease the number of false-positives, even at the cost of some additional false-negatives, which translates to increasing the indicator size (and in turn also increasing the update interval).
If, on the other hand, $(\missp-1)\cdot\fnp_t > \fpp_t$, we would like to do the converse.
We use the term $\sqrt{\frac{\fpp_t}{(\missp-1)\cdot\fnp_t}}$ as the step size (and direction) for updating the indicator size.
This step size implies the same factor for adjusting the update interval (for maintaining the budget constraint, as verified by line~\ref{alg:full_indicator_end}).
By this, we effectively distribute the required change of $\frac{\fpp}{(\missp-1)\cdot\fnp}$ equally across the indicator size (governing the behavior of false-positives) and the update interval (governing the behavior of false-negatives).
The combined effect brings us closer to having $\frac{\fpp}{(\missp-1)\cdot\fnp}=1$.
The algorithm may adjust the indicator size to ensure that it is within the allowed range, and then adjusts the update interval to satisfy Eq.~\ref{Eq:budget_const_full_ind}, and avoid violating the budget constraint.

\subsubsection{Delta-updates, no budget violation or non-minimal indicator size}
Lines~\ref{alg:delta_lambert_begin}-\ref{alg:delta_lambert_end} describe a case where either the budget constraint during the segment ending at $t$ was satisfied (i.e., $\bandwidth_t\leq B$), or the indicator size can still be reduced (i.e., $\abs{\ind_t} > \abs{\ind_{\min}}$).
We refer to this case as having the algorithm work in {\em Mode 2} (shaded blue).

Assume first that there is no budget violation.
By the discussion presented in Sec.~\ref{sec:alg:HL_description}, when sending delta-updates, and when there is no budget violation, it is advisable to send updates as fast as possible, i.e., using the minimal update interval $\minUpdateInterval$.
This approach also implies that there will be no (or very few) false-negatives, since we update the indicator with the shortest allowed interval, keeping it (almost) up to date.
For determining the indicator size, we note that the overall bandwidth available per request can be increased by a factor of $\frac{\budget}{\bandwidth_t}$.
It follows that we would like to utilize the entire budget to minimize the number of false-positives.
If, on the other hand, there is a budget violation, but $\indSizeT>\minIndSize$, this implies that we may remain in the delta regime, but will be forced to reduce the indicator size to stay within budget.

To determine the ratio by which we should adjust the indicator size, it is instructive to consider the effect of changing the indicator size $\abs{\ind_t}$ by some factor $\beta>0$.
Such an adjustment implies that every change in the cache will cause $\beta$ times more/less (depending on whether $\beta>1$ or not) changed bits in the indicator.
Furthermore, adjusting the indicator size by a factor of $\beta$ implies that specifying each index in the indicator would now requires $\log(\beta\cdot\indSizeT)$ bits instead of $\log \indSizeT$.
It follows that the overall number of bits sent for each change in the cache would increase/decrease by a factor of $\beta \cdot \frac{\log(\beta \cdot \indSizeT)}{\log\indSizeT}$.
We would like the overall change in the number of bits sent to be equal to $\frac{\budget}{\bandwidth_t}$, to match the budget.
Formally, we seek a new indicator size $\abs{\ind}$ s.t:
\begin{equation}
\frac{\indSize \log \indSize}{\indSizeT \log \indSizeT } = \frac{\budget}{\bandwidth_t},
\end{equation}
where we replace $\beta$ by $\frac{\indSize}{\indSizeT}$. The solution to this equation is obtained by using the Lambert $W$ function~\cite{lambert}, implying that the new indicator size should be set to
$\floor[\big]{
e^{W\left(
\frac{B \indSizeT\log\indSizeT}{\bandwidth_t}
\right)}
}$.\footnote{The Lambert $W$ function is the inverse of the function $f(w)=w e^{w}$. It can be used for solving the equation $x\ln x = a$ by substituting $y=\ln x$, resulting in $y e^{y}=a$, implying that $x=e^y=e^{W(a)}$.}
The algorithm then ensures that the best indicator size in this case falls within the allowed range.

\subsubsection{Delta-updates, budget violation, minimal indicator size}
The third and last case is when we use the minimal indicator size, but we still violate the budget.
We refer to this case as having the algorithm work in {\em Mode 3} (shaded green).

In such a case, the only way to ensure feasibility is to increase the update interval, as done in lines~\ref{alg:delta_min_violation_begin}-\ref{alg:delta_min_violation_end}, in order to satisfy Eq.~\ref{Eq:budget_const_full_ind}.
Such a scenario indeed occurs in practice (as we show in Sec.~\ref{sec:evaluation}), and handling this case ensures that the algorithm can return to the configurations covered by the previous two cases.  Mode 3 allows the algorithm to transcend to a considerably different state where we may prefer to send full indicators. Without it, \alg\ cannot leave the delta regime. 


\section{Performance Evaluation}
\label{sec:evaluation}
In this section, we present the results of our simulation study. 

\begin{figure*}[th]



\newcommand{\markerone}{*}
\newcommand{\markeronesize}{2pt}
\newcommand{\markertwo}{square}
\newcommand{\markertwosize}{3pt}

\newcommand{\cachesizeplot}[8]{
\begin{tikzpicture}
\begin{groupplot}[
    group style=
        {
        group size=3 by 4,
        xlabels at=edge bottom,
        ylabels at=edge left,
        horizontal sep=0.02\textwidth,
        vertical sep=0.06\textwidth,
        group name=plots,
        },
    major x tick style = transparent,
    ymajorgrids = true,
    width=0.5\columnwidth,
    ytick={1,1.5,2,2.5,3,3.5,4},
    yticklabel style={
        text width=0.035\textwidth,
        align=right,
        inner xsep=0pt,
        xshift=-0.014\textwidth
        },
    ylabel=cost,
    symbolic x coords={LRU,W-tLFU,FRD,Hyper},
    xticklabels={LRU,W-tLFU,FRD,Hyper},
    xticklabel style={
        font=\footnotesize,
        align=center,
        rotate=45,
        yshift=0.012\textwidth
        },
    xtick=data,
    ymin = 0.95,
    ymax= 4,
    legend style={
        legend columns=-1,
        at={(-0.645,1.6)},
        anchor=north,
        /tikz/every even column/.append style={column sep=0.3cm},
        },
    ]


\nextgroupplot[title=$4K$, ylabel={#1 $|$ cost}]
        \addplot[ybar=3pt,fill=cyan!50] table [ybar,x=policy,y=4KAlg]{#2};
        \addplot[only marks,mark=\markerone,mark size=\markeronesize,magenta] table [x=policy,y=4KBM]{#2};
        \addplot[only marks,mark=\markertwo,mark size=\markertwosize,black] table [x=policy,y=4KCF]{#2};
\nextgroupplot[title=$16K$,yticklabels=\empty]
        \addplot[ybar=3pt,fill=cyan!50] table [ybar,x=policy,y=16KAlg]{#2};
        \addplot[only marks,mark=\markerone,mark size=\markeronesize,magenta] table [x=policy,y=16KBM]{#2};
        \addplot[only marks,mark=\markertwo,mark size=\markertwosize,black] table [x=policy,y=16KCF]{#2};
\nextgroupplot[title=$64K$,yticklabels=\empty]
        \addplot[
            ybar=3pt,
            fill=cyan!50,
            ybar legend,
            legend image code/.code={%
               \draw (0cm,-0.1cm) rectangle (0.3cm,0.1cm);
            },
        ] table [ybar,x=policy,y=64KAlg]{#2};
        \addplot[only marks,mark=\markerone,mark size=\markeronesize,magenta] table [x=policy,y=64KBM]{#2};
        \addplot[only marks,mark=\markertwo,mark size=\markertwosize,black] table [x=policy,y=64KCF]{#2};

\legend{\alg,SC,CF}

\nextgroupplot[ylabel={#3 $|$ cost}]
        \addplot[ybar=3pt,fill=cyan!50] table [ybar,x=policy,y=4KAlg]{#4};
        \addplot[only marks,mark=\markerone,mark size=\markeronesize,magenta] table [x=policy,y=4KBM]{#4};
        \addplot[only marks,mark=\markertwo,mark size=\markertwosize,black] table [x=policy,y=4KCF]{#4};
\nextgroupplot[yticklabels=\empty]
        \addplot[ybar=3pt,fill=cyan!50] table [ybar,x=policy,y=16KAlg]{#4};
        \addplot[only marks,mark=\markerone,mark size=\markeronesize,magenta] table [x=policy,y=16KBM]{#4};
        \addplot[only marks,mark=\markertwo,mark size=\markertwosize,black] table [x=policy,y=16KCF]{#4};
\nextgroupplot[yticklabels=\empty]
        \addplot[
            ybar=3pt,
            fill=cyan!50,
            ybar legend,
            legend image code/.code={%
               \draw (0cm,-0.1cm) rectangle (0.3cm,0.1cm);
            },
        ] table [ybar,x=policy,y=64KAlg]{#4};
        \addplot[only marks,mark=\markerone,mark size=\markeronesize,magenta] table [x=policy,y=64KBM]{#4};
        \addplot[only marks,mark=\markertwo,mark size=\markertwosize,black] table [x=policy,y=64KCF]{#4};

\nextgroupplot[ylabel={#5 $|$ cost}]
        \addplot[ybar=3pt,fill=cyan!50] table [ybar,x=policy,y=4KAlg]{#6};
        \addplot[only marks,mark=\markerone,mark size=\markeronesize,magenta] table [x=policy,y=4KBM]{#6};
        \addplot[only marks,mark=\markertwo,mark size=\markertwosize,black] table [x=policy,y=4KCF]{#6};
\nextgroupplot[yticklabels=\empty]
        \addplot[ybar=3pt,fill=cyan!50] table [ybar,x=policy,y=16KAlg]{#6};
        \addplot[only marks,mark=\markerone,mark size=\markeronesize,magenta] table [x=policy,y=16KBM]{#6};
        \addplot[only marks,mark=\markertwo,mark size=\markertwosize,black] table [x=policy,y=16KCF]{#6};
\nextgroupplot[yticklabels=\empty]
        \addplot[
            ybar=3pt,
            fill=cyan!50,
            ybar legend,
            legend image code/.code={%
               \draw (0cm,-0.1cm) rectangle (0.3cm,0.1cm);
            },
        ] table [ybar,x=policy,y=64KAlg]{#6};
        \addplot[only marks,mark=\markerone,mark size=\markeronesize,magenta] table [x=policy,y=64KBM]{#6};
        \addplot[only marks,mark=\markertwo,mark size=\markertwosize,black] table [x=policy,y=64KCF]{#6};

\nextgroupplot[ylabel={#7 $|$ cost}]
        \addplot[ybar=3pt,fill=cyan!50] table [ybar,x=policy,y=4KAlg]{#8};
        \addplot[only marks,mark=\markerone,mark size=\markeronesize,magenta] table [x=policy,y=4KBM]{#8};
        \addplot[only marks,mark=\markertwo,mark size=\markertwosize,black] table [x=policy,y=4KCF]{#8};
\nextgroupplot[yticklabels=\empty]
        \addplot[ybar=3pt,fill=cyan!50] table [ybar,x=policy,y=16KAlg]{#8};
        \addplot[only marks,mark=\markerone,mark size=\markeronesize,magenta] table [x=policy,y=16KBM]{#8};
        \addplot[only marks,mark=\markertwo,mark size=\markertwosize,black] table [x=policy,y=16KCF]{#8};
\nextgroupplot[yticklabels=\empty]
        \addplot[
            ybar=3pt,
            fill=cyan!50,
            ybar legend,
            legend image code/.code={%
               \draw (0cm,-0.1cm) rectangle (0.3cm,0.1cm);
            },
        ] table [ybar,x=policy,y=64KAlg]{#8};
        \addplot[only marks,mark=\markerone,mark size=\markeronesize,magenta] table [x=policy,y=64KBM]{#8};
        \addplot[only marks,mark=\markertwo,mark size=\markertwosize,black] table [x=policy,y=64KCF]{#8};

\end{groupplot}
\end{tikzpicture}
} 



    \scalebox{0.8}{\cachesizeplot{F1}{F1.dat}{F2}{F2.dat}{Wiki1}{wiki.dat}{Wiki2}{wiki2.dat}}%
    \scalebox{0.8}{\cachesizeplot{Scarab}{scarab.dat}{P3}{P3.dat}{P6}{P6.dat}{P8}{P8.dat}}
    \caption{Access cost for \alg, the best feasible static configuration (SC), and always accessing the cache (CF).} 
    
    \label{fig:competitive:workload}
\end{figure*}
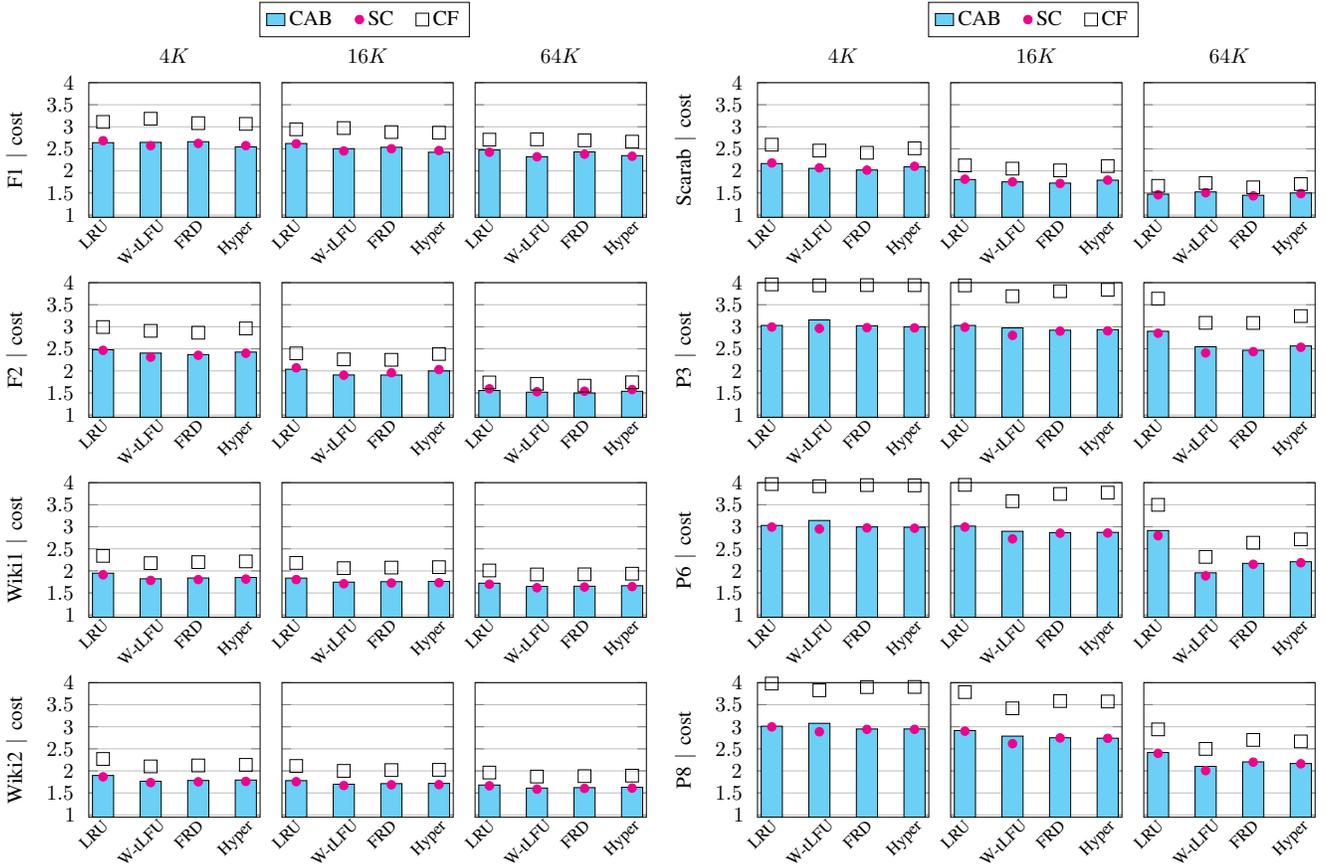

\textbf{Setup and system parameters:}
We focus on scenarios handled by general-purpose caching libraries such as Caffeine~\cite{caffeine}, 
Ristretto~\cite{ristretto}, Guava Cache~\cite{guava}, and the likes.
In particular, we use Caffeine for the evaluation of our proposed solution.\footnote{Caffeine is arguably the most popular Java libraries and is used in tens of large open-source projects such as Cassandra, Corfu, and Infinispan.}
The cache has a split {\em get/put} interface where {\em get} tests the cache, and {\em put} updates the cache. 
We extended Caffeine's simulator~\cite{caffeine}
to simulate the access cost with cache advertisements.
In our implementation, we issue a {\em get} request upon a positive indication, and we issue a {\em put} after handling the request regardless of indications.
Our advertisement mechanism uses the Orestes Bloom filters library~\cite{OrestesBf}. The cache maintains a four-bit Counting Bloom Filter (CBF)~\cite{CBF}. 
However, before sending an update, the 
CBF 
is compressed
to a simple Bloom Filter (BF)~\cite{Bloom}, where a bit in the BF is set iff the respective counter in the CBF is (strictly) positive.
The number of hash functions is optimized to minimize the false-positive rate~\cite{Survey12}.

We set $\minUpdateInterval = 10$ to minimize the  transmission overheads.  
We set $\minIndSize$ to $2.5 \cdot \cacheSize$ which suffices for a false-positive ratio of 30$\%$, and $\maxIndSize$ to $15 \cdot \cacheSize$ that implies a false-positive ratio of 0.07\%~\cite{Survey12}.
Our default budget is $\budget=20$, which sets $\maxUpdateInterval$ to $\frac{\maxIndSize}{B}=0.75\cdot C$. We assume a miss penalty $\missp=3$, which is typical for edge computing, where the delay from a cloud processing unit to the memory in the CDN is three times higher than the delay from the cache located at the edge~\cite{Octopus}.

\textbf{Benchmarks:}
The static configuration benchmark (SC) is the best static configuration across a grid of configurations satisfying the budget constraints in all segments. We consider a set of possible indicator sizes $\set{\minIndSize \cdot (1.1)^i | i = 0,\ldots,18} \subseteq [\minIndSize,\maxIndSize]$
(where $\minIndSize\cdot (1.1)^i \leq \maxIndSize$ implies in this case that $i \leq \floor{\frac{\log 6}{\log 1.1}}=18$),
and also the maximal indicator size $\maxIndSize$.
The update intervals considered are taken from the range $\set{\minUpdateInterval \cdot (1.15)^j | j = 0,\ldots,\frac{\log(\maxUpdateInterval/\minUpdateInterval)}{\log(1.15)}}$.
For example, for cache sizes $4K$, $16K$, and $64K$, the number of possible update intervals considered in the grid are 41, 51, and 61, respectively. We note that SC can only be determined in retrospect. 
We also evaluate the  \cachefirst\ (CF) policy that always accesses the cache (without indicators) to quantify the access cost reduction from using advertisements. 

{\bf Traces:}
We use the following real workloads, which are commonly used when evaluating caching systems:
\begin{inparaenum}[(i)]
\item {\em Scarab}:
A trace from Scarab Research, a personalized recommendation system for e-commerce sites~\cite{Scarab_and_Gradle_traces}.
\item {F1, F2}: Traces taken from a financial transaction processing system~\cite{Umass_traces}.
\item {\em P3, P6, P8}: Traces of disk accesses in Windows servers~\cite{ARC}.
\item {\em Wiki1, Wiki2}: Read requests to Wikipedia pages~\cite{WikiBench}.
\end{inparaenum}

{\bf Cache policies:}
We simulated the classic LRU policy, along with three highly competitive policies, including W-TinyLFU~\cite{TinyLFU} (also denoted by W-tLFU)),  FRD~\cite{FRD}, and Hyperbolic~\cite{Hyperbolic}. For completeness, we now very briefly outline these policies. 
The Least Recently Used (LRU) policy evicts the least recently accessed item. It assumes that recently accessed items would be accessed again.
The W-tLFU policy combines LRU with a frequency-based cache. Items are only admitted to a frequency-based cache if they are more frequent in a long history, which is represented as a CBF for space efficiency. 
FRD varies the time it retains admitted items according to their past access pattern. First-timers are admitted for a short duration, while previously encountered items are admitted for a longer duration. FRD uses extensive metadata about past accesses to distinguish first-timers from recurring items. 
Finally, Hyperbolic caching is an adaptive cache policy that changes its eviction policy according to the workload.

\subsection{Competitive Evaluation:}

{\bf \alg\ across workloads and policies:}
Our first experiment compares the access cost for the four cache policies when varying the cache size and the workload. 
Fig.~\ref{fig:competitive:workload} shows the results of these experiments. First, observe that increasing the cache size reduces the access costs, as expected by cache policies. Further, observe that for most traces, the differences between the cache policies are not very large, and are smaller than the differences between the CF and SC policy.
This implies that the potential benefit from advertising cache content may be higher than the benefit from changing the cache policy.
Finally, observe that the performance of \alg\ is very similar to that of the SC benchmark.
In some cases, (e.g., F1 4K LRU) \alg\ is even slightly better than the best static configuration.
Such a result implies that conditions change during the trace and that \alg\ manages to adjust itself according to these changes, thus reducing cost.
In other cases, \alg\ is slightly worse than the best configuration, but the difference is always small. 
\alg\ operates in {\em real time}, without prior knowledge of the system configuration, or the workload (other than knowing the cache size, the budget, and the bounds on the indicator sizes).

{\bf \alg\ across budgets and policies:}
Fig.~\ref{fig:competitive:budget} illustrates the results for varying budgets.
As expected, in all policies \alg\ and SC improve when the budget increases. As in our previous experiment, \alg\ matches the performance of SC regardless of budget. However, for W-tLFU \alg\ is not as good as SC for a small budget (10). The reason for this is that W-tLFU contains a very small and rapidly-changing Window cache, which forces \alg\ into short update intervals, which are bad for the larger and less dynamic Main cache (consuming 99\% of the cache space).  Alternatively, notice that  \alg\ is slightly better than SC for a budget of 40 across all policies,  implying that it manages to adapt to the changes within the trace. 


\newcommand{\markerone}{*}
\newcommand{\markeronesize}{2pt}
\newcommand{\markertwo}{square}
\newcommand{\markertwosize}{3pt}

\newcommand{\budgetplot}[5]{


\begin{tikzpicture}
\begin{groupplot}[
    group style=
        {
        group size=4 by 1,
        xlabels at=edge bottom,
        ylabels at=edge left,
        horizontal sep=0.02\textwidth,
        vertical sep=0.06\textwidth,
        group name=plots,
        },
    major x tick style = transparent,
    ymajorgrids = true,
    width=0.45\columnwidth,
    height=0.5\columnwidth,
    yticklabel style={
        align=right,
        inner xsep=0pt,
        xshift=-0.014\textwidth,
        },
    symbolic x coords={
        B1,
        B2,
        B3,
        B4,
        },
    enlarge x limits = abs{0.4},
    xticklabels={#2},
    xticklabel style={
        font=\footnotesize,
        align=center,
        },
    xtick=data,
    ymin = #4,
    ymax= #5,
    legend style={
        legend columns=-1,
        at={(-1.22,1.5)},
        anchor=north,
        /tikz/every even column/.append style={column sep=0.3cm},
        },
    ]

\nextgroupplot[title=LRU, ylabel={#3 $|$ cost}]
        \addplot[
            ybar=5pt,
            fill=cyan!50,
            ]
            table [
                ybar,
                x=budget,
                y=LRU-A,
                ]
                {#1};
        \addplot[only marks,mark=\markerone,mark size=\markeronesize,magenta] table [x=budget,y=LRU-SC]{#1};
        \addplot[only marks,mark=\markertwo,mark size=\markertwosize,black] table [x=budget,y=LRU-CF]{#1};

    \nextgroupplot[
        title=W-tLFU,
        yticklabels=\empty,
        x label style={
            at={(axis description cs:1.1,-0.15)},
            anchor=north
            },
        xlabel=Budget,
        ]
        \addplot[
            ybar=5pt,
            fill=cyan!50,
            ]
            table [
                ybar,
                x=budget,
                y=LFU-A,
                ]
                {#1};
        \addplot[only marks,mark=\markerone,mark size=\markeronesize,magenta] table [x=budget,y=LFU-SC]{#1};
        \addplot[only marks,mark=\markertwo,mark size=\markertwosize,black] table [x=budget,y=LFU-CF]{#1};

\nextgroupplot[title=FRD, yticklabels=\empty]
        \addplot[
            ybar=5pt,
            fill=cyan!50,
            ]
            table [
                ybar,
                x=budget,
                y=FRD-A,
                ]
                {#1};
        \addplot[only marks,mark=\markerone,mark size=\markeronesize,magenta] table [x=budget,y=FRD-SC]{#1};
        \addplot[only marks,mark=\markertwo,mark size=\markertwosize,black] table [x=budget,y=FRD-CF]{#1};

\nextgroupplot[title=Hyper, yticklabels=\empty]
        \addplot[
            ybar=5pt,
            fill=cyan!50,
            legend image code/.code={%
              \draw (0cm,-0.1cm) rectangle (0.3cm,0.1cm);
              },
            ]
            table [
                ybar,
                x=budget,
                y=Hyp-A,
                ]
                {#1};
        \addplot[only marks,mark=\markerone,mark size=\markeronesize,magenta] table [x=budget,y=Hyp-SC]{#1};
        \addplot[only marks,mark=\markertwo,mark size=\markertwosize,black] table [x=budget,y=Hyp-CF]{#1};


\legend{\alg,SC,CF}

\end{groupplot}
\end{tikzpicture}
} 

\begin{figure}[t!]
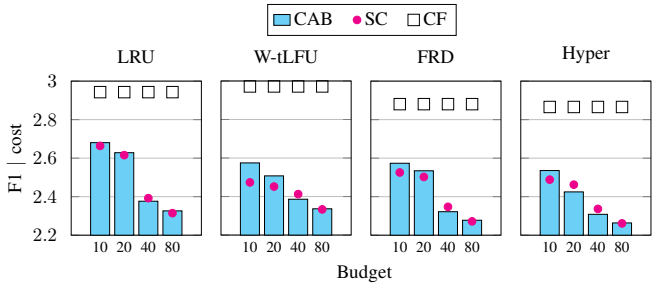

    \scalebox{0.72}{\budgetplot{F1.budget.dat}{10,20,40,80}{F1}{2.2}{3}}
    \caption{Access cost for \alg, the best feasible static configuration (SC), and always accessing the cache (CF), for distinct budgets and 4 policies. The workload is F1 and the cache size is $\cacheSize=16$K.}
    \label{fig:competitive:budget}
\end{figure}

\subsection{\alg\ Under the Hood:}
To highlight the performance and behavior of \alg\ in dynamic settings, where the workload changes, we run several traces one after another.
Since each workload's characteristics are slightly different, such an experiment allows us to follow the dynamic change of configuration performed by \alg\ and the system behavior as it interacts with these changes.

\begin{figure}[t!]
    \centering
    \scalebox{1}{\input{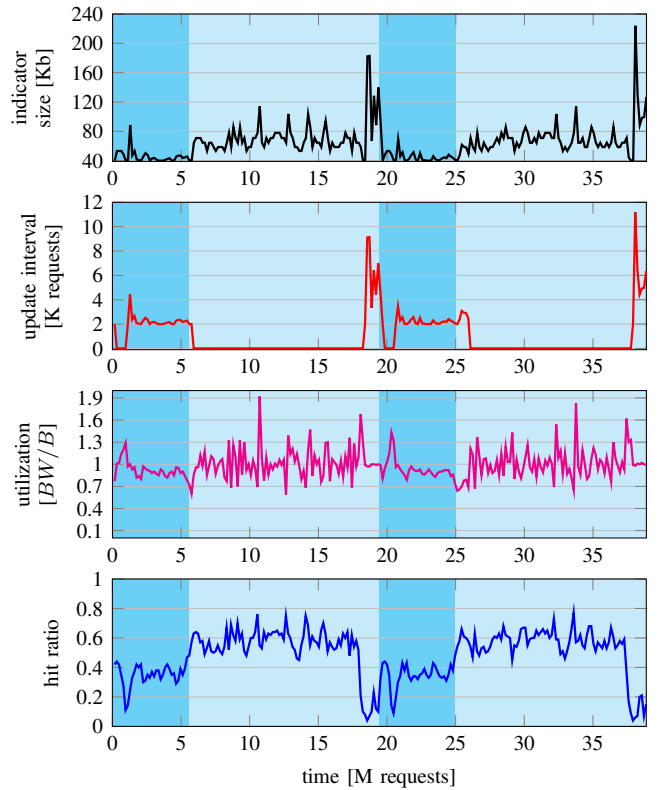}}
    \caption{The dynamics of \alg\ on a concatenated trace F1$\rightarrow$F2$\rightarrow$F1$\rightarrow$F2. F1 requests are dark shaded, and F2 requests are light shaded.
    The cache size is $16K$, the policy is LRU, and the budget is 20 bits/request.}
    \label{fig:concat_trace}
    \vspace{-0.2cm}
\end{figure}

Fig.~\ref{fig:concat_trace}
follows
the execution of \alg\ for a combination of traces F1 (dark shaded area) and F2 (light shaded area), concatenated as F1$\rightarrow$F2$\rightarrow$F1$\rightarrow$F2. Note that both F1 and F2 are typical to the same application (financial transactions), and therefore such a concatenated workload may indeed happen in practice.
The total request count is $\sim$40M. The experiment is performed using $\cacheSize=16$K, the LRU policy, $\missp=3$, and a budget of 20 bits/req. 
As in earlier experiments, we take $\minUpdateInterval=10$, and $\alpha=10$. This implies $T=160$K, resulting in $\sim$250 configuration segments during the entire simulation. Also, we set $\minIndSize=2.5 \cdot \cacheSize=40$K, and $\maxIndSize=15 \cdot \cacheSize=240$K.

The top subfigure shows the evolution of the indicator size. The second topmost subfigure shows the evolution of the update interval. These two figures provide a glimpse into the evolution of the configurations used by \alg.
The bottom subfigure shows the evolution of the hit ratio of the cache. 
We stress that the hit ratio captures the ratio of requests that are actually found in the cache, and is therefore an artifact of the settings (workload, cache size, and cache policy). The advertisement policy doesn't impact the hit ratio. However, \alg's advertisement policy implicitly {\em reacts} to the hit ratio.
The second bottom-most subfigure shows the bandwidth utilization (normalized to the bandwidth budget). Notice that the utilization is calculated before network policing; hence \alg\ may try to exceed the bandwidth (but the network policing prevents that). 

These figures show the effect of the algorithm's choices and the properties of the workload. We now turn to explain and describe the algorithm's performance along time $t$ (measured by the request counts). 
\begin{itemize}[leftmargin=12pt]
\item $t \in [0, 1\Meg]$: Soon after the beginning, \alg\ identifies that the cache uses delta-updates, and thus sets $\updateintervalT=\minUpdateInterval$, while adjusting the indicator size so as to comply with the budget constraint  (Mode 2).
\item $ t \sim 1\Meg$:
The hit ratio (bottom subfigure) sharply decreases. This translates to a substantial change in the cached content, resulting in much larger delta updates that violate the budget constraint (second-bottom subfigure). Note that \alg\ cannot exceed the budget, as its advertisements are dropped once the budget is exhausted (leading to additional errors). 
The algorithm fails to comply with the budget constraint with a minimal update interval, even when shrinking the indicator size to $\minIndSize$. Hence, \alg\ increases the update interval, 
to ensure adhering to the budget restriction (Mode 3). 
\item $t \in [1\Meg, 5.5\Meg]$: \alg\ constantly works in Mode 1, mostly sending full indicators (as it is cheaper than sending the mere changes). During this time, the algorithm constantly satisfies the budget constraint, and merely makes small adjustments to the indicator size, and the update interval, to optimize its usage of bandwidth while balancing the extra costs incurred by false-positives and false-negatives.
\item $t \in [5.5\Meg, 18\Meg]$: the F1 trace ends, and the F2 trace begins. The hit ratio
increases, thus allowing for using a much larger indicator, and more frequent updates. The algorithm hence switches to Mode 2, and persistently uses the minimal update interval.
When
\alg\ occasionally violates the budget constraint,
it shrinks the indicator again, to stay within limits.
\item $t \sim 18\Meg$: A sudden drop in the hit ratio does not allow the algorithm to keep a minimal update interval anymore, even when the indicator size is $\minIndSize$. Hence, \alg\ switches to mode 3, and significantly increases the update interval to satisfy the budget constraint.
\item $t \in [18\Meg, 20\Meg]$: The algorithm works in Mode 1, constantly sending full indicators, and adjusting the indicator size and the update interval to remain within budget.
The general increase in hit ratio allows the algorithm to settle for smaller indicator size and update intervals.
\item $t \in [20\Meg, 25.5\Meg]$: The algorithm again behaves as it did when initially handling F1. 
\item $t \in [25.5\Meg, 40\Meg]$: The F2 trace arrives, and the algorithm exhibits similar behavior when transitioning again to handling F2. However, a closer look shows that the indicator size and the update interval slightly differ from those chosen by the algorithm in the previous run of F2, which occurred in the interval $t \in [5.5\Meg, 18\Meg]$. These changes further exemplify how the algorithm adapts even to minor changes stemmed from differences in the cache's content at the beginning of the different runs of F2.
\end{itemize}

\section{Discussion and Conclusions}
Many systems use cache advertisements, which highlight the need for {\em efficient} cache advertisement strategies.
Yet, prior to our work, the literature lacked a rigorous method to configure the advertisement strategy.
Therefore, system designers turn to design decisions based on rules-of-thumb and ad-hoc benchmarks of typical workloads. Additionally, these approaches are mostly limited to selecting a static advertisement policy, which means that changes in system and workload parameters might degrade the quality of their choice. 

Our work surveys the possible modes of operations that an advertisement policy can utilize.
We empirically show that there is no one-size-fits-all policy and that advertisement policies' performance depends on the cache policy, the cache size, and the workload. Worse yet, static policies are ill-suited for adaptive cache policies that change their behavior during run-time~\cite{Adaptive_Sw_Cache,Hyperbolic,ARC}.  

We designed the novel \alg\ algorithm that adjusts the advertisement policy according to the current conditions.
\alg\ reaches its decision by monitoring its bandwidth footprint, false-positive, and false-negative errors. It is indifferent to the cache size, the workload, and the cache policy (beyond their indirect effect on the false-positive and false-negative rates).

We performed an extensive evaluation that uses eight real workloads and tested the classic LRU policy and three other leading cache management policies.
We performed our work under a strict network model that drops messages if transmitting them would violate the bandwidth budget. 
Under these conditions, \alg\ exhibits an overall cost comparable (and sometimes superior) to the best static advertisement policy for all cache sizes, workloads, and cache policies.
\alg\ is a game-changer as developers no longer need to optimize their advertisement policy manually.
Instead, they can use \alg\ to optimize the advertisement strategy, shorten development time, and attain good performance in a variety of scenarios and system configurations.
More so, \alg\ solves problems  that were encountered by many  works~\cite{b1,b2,b3,b4,b5,fpr_fnr_in_dist_replicas,summary_cache, updateIntervalWebCache05, uIntervalInMANET, uIntervalSimsInCCN}, and solved only in an ad-hoc manner. 

Next,
we study and discuss the dynamic behavior of \alg,
varying 
its advertisement strategy according to the conditions, and effectively
transitioning
between the delta and full-indicator updates regimes.
To the best of our knowledge, \alg\ is the first to combine these options seamlessly. 
We note that \alg\ changes its transmission policy quite often during the system's lifetime, which incurs computation costs at the cache. E.g.,
a new indicator is required
every time we change the indicator size. While we do not evaluate the CPU usage of the cache, we sized the reconfiguration intervals to be ten times the cache size. Thus, the amortized cost of computing a new indicator is one indicator operation per 10 cache accesses. 
We believe that such a configuration makes the additional overheads manageable (if not negligible) since current indicators, such as Bloom filter implementations~\cite{OrestesBf,caffeine}, reach over 20 million ops per second on a single thread and are embarrassingly parallel.

Another takeaway from our research is that false-negatives should not be neglected when advertising cached content.
Looking into the future, we plan to develop access strategies that cope well with false-negatives, and then use such strategies alongside \alg\ on a distributed network. Such work may borrow ideas from~\cite{fpr_fnr_in_dist_replicas,Accs_Strategies_Infocom} in developing improved solutions for distributed cache access with advertisements.

\bibliographystyle{IEEEtran}
\bibliography{Refs}

\begin{thebibliography}{10}
\providecommand{\url}[1]{#1}
\csname url@samestyle\endcsname
\providecommand{\newblock}{\relax}
\providecommand{\bibinfo}[2]{#2}
\providecommand{\BIBentrySTDinterwordspacing}{\spaceskip=0pt\relax}
\providecommand{\BIBentryALTinterwordstretchfactor}{4}
\providecommand{\BIBentryALTinterwordspacing}{\spaceskip=\fontdimen2\font plus
\BIBentryALTinterwordstretchfactor\fontdimen3\font minus
  \fontdimen4\font\relax}
\providecommand{\BIBforeignlanguage}[2]{{%
\expandafter\ifx\csname l@#1\endcsname\relax
\typeout{** WARNING: IEEEtran.bst: No hyphenation pattern has been}%
\typeout{** loaded for the language `#1'. Using the pattern for}%
\typeout{** the default language instead.}%
\else
\language=\csname l@#1\endcsname
\fi
#2}}
\providecommand{\BIBdecl}{\relax}
\BIBdecl

\bibitem{BloomParadox}
O.~Rottenstreich and I.~Keslassy, ``The bloom paradox: When not to use a bloom
  filter,'' \emph{{IEEE/ACM} Trans. Netw.}, vol.~23, no.~3, pp. 703--716, 2015.

\bibitem{Joint_opt}
X.~Guo, T.~Wang, and S.~Wang, ``Joint optimization of caching and routing
  strategies in content delivery networks: A big data case,'' in \emph{IEEE
  ICC}, 2019.

\bibitem{CDN_theory_Vs_practice}
B.~M. Maggs and R.~K. Sitaraman, ``Algorithmic nuggets in content delivery,''
  \emph{ACM SIGCOMM Computer Communication Review}, vol.~45, no.~3, pp. 52--66,
  2015.

\bibitem{uIntervalInMANET}
I.-W. Ting and Y.-K. Chang, ``Improved group-based cooperative caching scheme
  for mobile ad hoc networks,'' \emph{J. Parallel. and Distrib. Comp.},
  vol.~73, no.~5, pp. 595--607, 2013.

\bibitem{Digest_in_Manet}
T.~Le, Y.~Lu, and M.~Gerla, ``Social caching and content retrieval in
  disruption tolerant networks (dtns),'' in \emph{International Conference on
  Computing, Networking and Communications (ICNC)}.\hskip 1em plus 0.5em minus
  0.4em\relax IEEE, 2015, pp. 905--910.

\bibitem{CDN_OceanStore}
J.~Kubiatowicz, D.~Bindel, Y.~Chen, S.~Czerwinski, P.~Eaton, D.~Geels,
  R.~Gummadi, S.~Rhea, H.~Weatherspoon, W.~Weimer, C.~Wells, and B.~Zhao,
  ``Oceanstore: An architecture for global-scale persistent storage,''
  \emph{SIGPLAN Not.}, vol.~35, no.~11, pp. 190--201, 2000.

\bibitem{CDN_AdaptSize}
D.~S. Berger, R.~K. Sitaraman, and M.~Harchol-Balter, ``Adaptsize:
  Orchestrating the hot object memory cache in a content delivery network,'' in
  \emph{NSDI}, 2017, pp. 483--498.

\bibitem{ICN}
M.~Bilal and S.~G. Kang, ``A cache management scheme for efficient content
  eviction and replication in cache networks,'' \emph{IEEE Access}, vol.~5, pp.
  1692--1701, 2017.

\bibitem{ICN2}
I.~Psaras, W.~K. Chai, and G.~Pavlou, ``Probabilistic in-network caching for
  information-centric networks,'' in \emph{ICN}, 2012, pp. 55--60.

\bibitem{summary_cache}
L.~Fan, P.~Cao, J.~Almeida, and A.~Z. Broder, ``Summary cache: a scalable
  wide-area web cache sharing protocol,'' \emph{IEEE/ACM transactions on
  networking}, vol.~8, no.~3, pp. 281--293, 2000.

\bibitem{Bloom}
B.~H. Bloom, ``Space/time trade-offs in hash coding with allowable errors,''
  \emph{Commun. {ACM}}, vol.~13, no.~7, pp. 422--426, 1970.

\bibitem{CBF}
F.~Bonomi, M.~Mitzenmacher, R.~Panigrahy, S.~Singh, and G.~Varghese, ``An
  improved construction for counting bloom filters,'' in \emph{{ESA}}, 2006,
  pp. 684--695.

\bibitem{TinySet}
G.~Einziger and R.~Friedman, ``Tinyset: An access efficient self adjusting
  bloom filter construction,'' \emph{IEEE/ACM Trans. Netw.}, vol.~25, no.~4,
  pp. 2295--2307, 2017.

\bibitem{Survey12}
S.~Tarkoma, C.~E. Rothenberg, and E.~Lagerspetz, ``Theory and practice of bloom
  filters for distributed systems,'' \emph{IEEE Comm. Surv. \& Tut.}, vol.~14,
  no.~1, pp. 131--155, 2012.

\bibitem{Survey18}
L.~Luo, D.~Guo, R.~T. Ma, O.~Rottenstreich, and X.~Luo, ``Optimizing bloom
  filter: Challenges, solutions, and comparisons,'' \emph{IEEE Communications
  Surveys \& Tutorials}, vol.~21, no.~2, pp. 1912--1949, 2018.

\bibitem{TinyTable}
G.~Einziger and R.~Friedman, ``Counting with tinytable: Every bit counts!'' in
  \emph{{ICDCN}}, 2016, p.~27.

\bibitem{AccessEfficientBF}
Y.~Kanizo, D.~Hay, and I.~Keslassy, ``Access-efficient balanced bloom
  filters,'' \emph{Comput. Comm.}, vol.~36, no.~4, pp. 373--385, 2013.

\bibitem{updateIntervalWebCache05}
W.~Shi and Y.~Mao, ``Performance evaluation of peer-to-peer web caching
  systems,'' \emph{J. of Sys. and Soft.}, vol.~79, no.~5, pp. 714--726, 2006.

\bibitem{uIntervalSimsInCCN}
M.~Tortelli, L.~A. Grieco, and G.~Boggia, ``{CCN} forwarding engine based on
  bloom filters,'' in \emph{CFI}, 2012, pp. 13--14.

\bibitem{SquidFAQ}
\BIBentryALTinterwordspacing
{Squid Cache}, ``Squid-cache wiki.'' [Online]. Available:
  \url{https://wiki.squid-cache.org/SquidFaq/CacheDigests#Would_it_be_possible_to_stagger_the_timings_when_cache_digests_are_retrieved_from_peers.3F}
\BIBentrySTDinterwordspacing

\bibitem{SquidSpec}
\BIBentryALTinterwordspacing
------, ``Squid digest spec, v5.'' [Online]. Available:
  \url{http://www.squid-cache.org/CacheDigest/cache-digest-v5.txt}
\BIBentrySTDinterwordspacing

\bibitem{indicators_in_NDN19}
R.~Hou, L.~Zhang, T.~Wu, T.~Mao, and J.~Luo, ``Bloom-filter-based request node
  collaboration caching for named data networking,'' \emph{Cluster Computing},
  vol.~22, no.~3, pp. 6681--6692, 2019.

\bibitem{ICN_survey_15}
M.~Zhang \emph{et~al.}, ``A survey of caching mechanisms in information-centric
  networking,'' \emph{IEEE Communications Surveys \& Tutorials}, vol.~17,
  no.~3, pp. 1473--1499, 2015.

\bibitem{ICN_survey13}
G.~Zhang, Y.~Li, and T.~Lin, ``Caching in information centric networking: A
  survey,'' \emph{Comp. Net.}, vol.~57, no.~16, pp. 3128--3141, 2013.

\bibitem{Dynamic_BF}
D.~Guo, J.~Wu, H.~Chen, Y.~Yuan, and X.~Luo, ``The dynamic bloom filters,''
  \emph{IEEE Trans on Knowl. and Data Eng.}, vol.~22, no.~1, pp. 120--133,
  2009.

\bibitem{Accs_Strategies_Infocom}
I.~Cohen, G.~Einziger, R.~Friedman, and G.~Scalosub, ``Access strategies for
  network caching,'' in \emph{IEEE INFOCOM}, 2019, pp. 28--36.

\bibitem{chen2020sequential}
L.~Chen, A.~Giovanidis, W.~Wang, and L.~Shan, ``Sequential resource access:
  Theory and algorithm,'' \emph{INFOCOM}, 2021.

\bibitem{FN_aware}
I.~Cohen, G.~Einziger, and G.~Scalosub, ``On the power of false negative
  awareness in indicator-based caching systems,'' in \emph{IEEE ICDCDS}, 2021.

\bibitem{CompressedBF}
M.~Mitzenmacher, ``Compressed bloom filters,'' \emph{IEEE/ACM Trans. Netw.},
  vol.~10, no.~5, pp. 604--612, 2002.

\bibitem{FPfree_Ori}
S.~Z. Kiss, {\'E}.~Hosszu, J.~Tapolcai, L.~R{\'o}nyai, and O.~Rottenstreich,
  ``Bloom filter with a false positive free zone,'' in \emph{IEEE INFOCOM},
  2018, pp. 1412--1420.

\bibitem{HBA_journal}
Y.~Zhu, H.~Jiang, J.~Wang, and F.~Xian, ``{HBA}: Distributed metadata
  management for large cluster-based storage systems,'' \emph{IEEE Trans.
  Parallel Distrib. Syst.}, vol. 147, pp. 204--220, 2018.

\bibitem{fpr_fnr_in_dist_replicas}
Y.~Zhu and H.~Jiang, ``False rate analysis of bloom filter replicas in
  distributed systems,'' in \emph{{ICPP}}, 2006, pp. 255--262.

\bibitem{b1}
{Hong Tang} and {Tao Yang}, ``An efficient data location protocol for
  self.organizing storage clusters,'' in \emph{ACM/IEEE Supercomputing}, 2003,
  pp. 53--53.

\bibitem{b5}
H.~Cai and J.~Wang, ``Foreseer: A novel, locality-aware peer-to-peer system
  architecture for keyword searches,'' in \emph{ACM/IFIP/USENIX Middleware},
  2004, p. 38–58.

\bibitem{LRUVsFifoRoyOhad}
O.~Eytan, D.~Harnik, E.~Ofer, R.~Friedman, and R.~Kat, ``It's time to revisit
  {LRU} vs. {FIFO},'' in \emph{HotStorage}, 2020.

\bibitem{Digest}
A.~Rousskov and D.~Wessels, ``Cache digests,'' \emph{Comp. Net. and ISDN Sys.},
  vol.~30, no. 22-23, pp. 2155--2168, 1998.

\bibitem{Adaptive_Sw_Cache}
G.~Einziger, O.~Eytan, R.~Friedman, and B.~Manes, ``Adaptive software cache
  management,'' in \emph{ACM Middleware}, 2018, pp. 94--106.

\bibitem{ARC}
N.~Megiddo and D.~S. Modha, ``Arc: A self-tuning, low overhead replacement
  cache.'' in \emph{Fast}, no. 2003, 2003.

\bibitem{Hyperbolic}
A.~Blankstein, S.~Sen, and M.~J. Freedman, ``Hyperbolic caching: Flexible
  caching for web applications,'' in \emph{USENIX ATC}, 2017, pp. 499--511.

\bibitem{lambert}
R.~M. Corless, G.~H. Gonnet, D.~E.~G. Hare, D.~J. Jeffrey, and D.~E. Knuth,
  ``On the {Lambert} {$W$} function,'' \emph{Adv. Comput. Math.}, vol.~5,
  no.~1, pp. 329--359, 1996.

\bibitem{caffeine}
\BIBentryALTinterwordspacing
B.~Manes, ``Caffeine: A high performance caching library for java.'' [Online].
  Available: \url{https://github.com/ben-manes/caffeine}
\BIBentrySTDinterwordspacing

\bibitem{ristretto}
\BIBentryALTinterwordspacing
{Dgraph Labs, Inc.}, ``Ristretto: A high performance memory-bound go cache.''
  [Online]. Available: \url{https://github.com/dgraph-io/ristretto}
\BIBentrySTDinterwordspacing

\bibitem{guava}
\BIBentryALTinterwordspacing
Google, ``Guava: Google core libraries for java.'' [Online]. Available:
  \url{https://github.com/google/guava}
\BIBentrySTDinterwordspacing

\bibitem{OrestesBf}
\BIBentryALTinterwordspacing
{Baqend GmbH}, ``Orestes: Bloom filter library for java.'' [Online]. Available:
  \url{https://github.com/Baqend/Orestes-Bloomfilter}
\BIBentrySTDinterwordspacing

\bibitem{Octopus}
T.~X. Tran and D.~Pompili, ``Octopus: A cooperative hierarchical caching
  strategy for cloud radio access networks,'' in \emph{IEEE MASS}, 2016, pp.
  154--162.

\bibitem{Scarab_and_Gradle_traces}
\BIBentryALTinterwordspacing
``Caffeine's simulator cache traces.'' [Online]. Available:
  \url{https://github.com/ben-manes/caffeine/tree/master/simulator/src/main/resources/com/github/benmanes/caffeine/cache/simulator/parser}
\BIBentrySTDinterwordspacing

\bibitem{Umass_traces}
\BIBentryALTinterwordspacing
M.~Liberatore and P.~Shenoy, ``Umass trace repository,'' 2016. [Online].
  Available: \url{http://traces.cs.umass.edu/}
\BIBentrySTDinterwordspacing

\bibitem{WikiBench}
G.~Urdaneta, G.~Pierre, and M.~van Steen, ``Wikipedia workload analysis for
  decentralized hosting,'' \emph{Comp. Net.}, vol.~53, no.~11, pp. 1830--1845,
  2009.

\bibitem{TinyLFU}
G.~Einziger, R.~Friedman, and B.~Manes, ``Tinylfu: {A} highly efficient cache
  admission policy,'' \emph{{TOS}}, vol.~13, no.~4, pp. 35:1--35:31, 2017.

\bibitem{FRD}
S.~Park and C.~Park, ``{FRD}: A filtering based buffer cache algorithm that
  considers both frequency and reuse distance,'' in \emph{MSST}, 2017.

\end{thebibliography}

\end{document}